\documentclass[aps,prd,showpacs,floatfix,preprint,preprintnumbers,a4paper]{revtex4-1}
\usepackage{amsfonts,amsmath,units,wasysym,epsfig,graphicx,verbatim,color,subfigure,graphicx,bm,mathrsfs,lipsum,hyperref}
\usepackage[bottom]{footmisc}
\usepackage[normalem]{ulem}  % \sout{old text} for strikeout

\begin{document}

\newcommand{\IUCAA}{Inter-University Centre for Astronomy and
  Astrophysics, Post Bag 4, Ganeshkhind, Pune 411 007, India}
\newcommand{\TIFR}{Department of Nuclear and Atomic Physics, Tata Institute of Fundamental Research, Mumbai 400005, India}  
\newcommand{\WSU}{Department of Physics \& Astronomy, Washington State University, 1245 Webster, Pullman, WA 99164-2814, U.S.A}
\newcommand{\INFN}{INFN Sezione di Ferrara, Via Saragat 1, I-44100 Ferrara, Italy}  
  
\title{Role of crustal physics in the tidal deformation of a neutron star}

\author{Bhaskar Biswas $^{\rm 1}$, Rana Nandi $^{\rm 2}$, Prasanta Char $^{\rm 3, 1}$, Sukanta Bose $^{\rm 1, 4}$}
\affiliation{$^{\rm 1}$ \IUCAA,$^{\rm 2}$ \TIFR,$^{\rm 3}$ \INFN,$^{\rm 4}$\WSU}

\begin{abstract}

In the late inspiral phase, gravitational waves from binary neutron star mergers carry the imprint of the equation of state due to the tidally deformed structure of the components. If the stars contain solid crusts, then their shear modulus can affect the deformability of the star and, thereby, modify the emitted signal. Here, we investigate the effect of realistic equations of state (EOSs) of the crustal matter, with a realistic model for the shear modulus of the stellar crust in a fully general relativistic framework. This allows us to systematically study the deviations that are expected from fluid models. In particular, we use unified EOSs, both relativistic and non-relativistic, in our calculations. 
We find that realistic EOSs of crusts cause a small correction, of $\sim 1\%$, in the second Love number. This correction will likely be subdominant to the statistical error expected in LIGO-Virgo observations at their respective advanced design sensitivities, but rival that error in third generation detectors.
%and therefore, not significant to constrain the EOS of neutron stars via astrophysical observations.
For completeness, we also study the effect of crustal shear on the magnetic-type Love number and find it to be much smaller.

\end{abstract}

\preprint{LIGO-P1900071}

\maketitle

\section{Introduction}

The detection of gravitational waves (GWs) from the binary neutron star (BNS) merger event GW170817 has ushered in a new probe for constraining the equation of state (EOS) of neutron stars~\cite{Abbott2017,Abbott2019,AbbottGRB}.
%CHECK: Also cite param and EOS papers on gw170817.
%GW astronomy . 
%At later stages of the 
Post-Newtonian theory predicts that  the inspiral signal from BNSs will carry an imprint of the EOS of neutron stars ~\cite{Flanagan2008,Hinderer2008,Binnington2009,Damour2009}.
% CITE damour ; poisson; hinderer papers
This result has been vetted by Numerical Relativity simulations as well.
% cite nr papers
%inspiral, there should be some imprint of internal structure of the neutron stars (NS) in the GW signal due to their combined tidal interactions \cite{Flanagan2008,Hinderer2008,Binnington2009,Damour2009}. 
One can use this imprint to study the properties of dense matter far from the nuclear saturation density with an event having significantly high signal-to-noise ratio in the future generation of detectors \cite{Hinderer2010}. 
%CITE Del Pozzo etc.
Neutron stars are also believed to have a solid crust as their outermost layers  \cite{Chamel2008}. The effect of the crust is a crucial ingredient for probing nuclear physics through GWs. The pioneering work of Carter and collaborators that introduced the theory of elastic solids in general relativity (GR)~\cite{Carter1972,Carter1973} has paved the way for studying the effect of NS crust in a consistent relativistic framework. Many of the studies concerning the NS perturbation incorporating crust elasticity have used the Cowling approximation \cite{Yoshida2002,Flores2017}. By contrast, there exist only a few studies that have accounted for full GR effects in the analysis \cite{Schumaker1983,Finn1990}. 

In one of the first attempts in this direction, Penner et al. tried to extract tidal information by employing an elastic crust~\cite{Penner2011} 
and followed it up  in another work to study the crustal failure  during BNS inspiral ~\cite{Penner2012}. However, they used mostly modest details of dense matter and rudimentary crust models. Therefore, the tidal behavior of different crust models inspired from various realistic nuclear interactions has not been very clear from their results. At this point, it should be mentioned that there have been other studies that investigate the tidal deformability of solid quark star using similar GR perturbative framework \cite{Lau2017,Lau2018}. The point of interest in them is mostly the phenomenology of a solid core of a star forming due to a deconfinement phase-transition at the center. In our case, we solely focus on the effect of a solid crust encapsulating a fluid core. 

The elastic properties of NS crust strongly depend on the composition of matter across a range of sub-nuclear densities. In the outer crust, the nuclei are arranged in the form of a body-centered-cubic (BCC) lattice that is embedded in a noninteracting and degenerate electron gas~\cite{Baym:1971pw,Nandi:2010fp}. As the density increases with depth, the neutron-drip point ($ \sim 3 \times 10^{-4}$ fm$^{-3}$) is reached, which signals the beginning of the inner crust. In this region, the neutron-rich nuclei are arranged as a lattice immersed in inter-penetrating gas of free neutrons and electrons~\cite{Baym:1971ax, Negele1973,Haensel2001,Nandi2011}. The shear modulus is also higher in the inner crust. Thus, the inner crust contributes the most towards the tidal response due to shear. This region extends till the crust-core transition density ($\sim 8 \times 10^{-4}$ fm$^{-3}$). Complex structures (e.g., rod, slab, bubble, etc. -- collectively known as nuclear pasta) are expected to occur in the inner crust as the matter gradually changes from crystalline to homogeneous phase, with increasing density~\cite{1983PhRvL..50.2066R,1984PThPh..71..320H,2016PhRvC..94b5806N,2018ApJ...852..135N} . 
% Laboratory experiments are yet to fully constrain the properties of matter at this density. 
Beneath the inner crust, the outer core starts, with uniform nuclear matter. As the density grows even higher, one reaches the inner core that can have superfluid neutrons and, perhaps, even exotic matter like strange baryons, meson condensates, quark matter, etc.~\cite{Glendenning2001,Page2006}. The characteristics of the matter in the core is highly speculative. Additionally, the EOS of the crust is qualitatively different and represents different physical conditions than that in the core. Therefore, one has to rely on proper matching of both EOSs at the crust-core interface. This is very crucial as it has been found that a proper thermodynamically consistent matching is required to avoid large uncertainties on the macroscopic properties of the star \cite{Fortin2016}.
Even then there might be some ambiguity due to different choices of crust-core transition density.
The best way is to use unified EOS models where the EOSs of the crust and the core are calculated within the same underlying theory. Hence, we employ unified EOSs in our work.

The main aim of this paper is to provide a comprehensive picture of the interplay between the perturbative response of the elastic crust of a NS and the nuclear physics of the constituents of the crustal matter using several unified EOS models. We have reworked the formalism of Penner et al.~\cite{Penner2011}  using the analysis of the perturbed quantities from Finn~\cite{Finn1990}. We find that realistic EOSs of crusts, with a non-zero shear modulus, cause a small correction, of $\sim 1\%$, in the second Love number. 
This correction will likely be subdominant to the statistical error expected in LIGO-Virgo observations at their respective advanced design sensitivities, but rival that error in third generation detectors.

The paper is organized as follows. In Sec.~\ref{sec:setup}, we discuss the formalism for tidal deformation and derivation of the Love numbers in elastic relativistic stars. Thereafter, we present an overview of the EOS used in our calculation in Sec.~\ref{EOS}. We discuss our results in Sec.~\ref{sec:results} and summarize in Sec.~\ref{sec:conclusion}. Throughout our analysis we have assumed $c=G=1$. 

\section{Analytical set up}
\label{sec:setup}

In this section, we present the analytical formulation of our work. Our focus is to calculate the tidal deformation of neutron stars with a solid crust. A solid crust supports shear stress and as a result two different types of pulsation mode arise. The odd parity modes are called torsional modes which creates twist in the star. These are first discussed in 1983 by Schumaker and Thorne~\cite{Schumaker1983}. Then in 1990 Finn~\cite{Finn1990} presented a new set of even parity type modes for a solid star. Since our focus is to calculate the tidal deformation of a neutron star we only need static perturbation equations which are basically zero frequency modes of pulsation problem. The set of static polar perturbation equations for a solid star are first given by Penner et al~\cite{Penner2011}. However, we found some inconsistencies in their equations, in particular, they don't match with the zero frequency limit of pulsation equations given by Finn~\cite{Finn1990}. Recently, Lau et al.~\cite{Lau2018} has also pointed out same sort of inconsistencies. Therefore, we re-derive all the equations in this work. Throughout the paper, we follow the notation of Thorne and Campolattaro~\cite{Thorne1967} as was also adopted by Finn~\cite{Finn1990}, so that we can easily verify our equations with that of  Finn's at zero frequency limit. 

Our aim is to quantify the effect of elasticity of crust in the tidal deformabilty of neutron stars. For this we  first calculate the background of the star by solving the standard Tolman-Oppenheimer-Volkoff (TOV) equation. 
In the next step we consider a static linear perturbation of this background model which takes into account elastic crust. To study the linear perturbation we  expand each components of fluid displacement vector and perturbed metric in terms of spherical harmonics. Each spherical harmonics is characterized by $l$, $m$ and parity, which can be either even $(-1)^{l}$ or odd $(-1)^{l+1}$. Under small amplitude motions these two parity decouple from each other and, hence, can be treated separately. Here we consider both cases individually and compute the deviation in Love number.

 \subsection{Background problem}

The equilibrium of a static, spherically symmetric relativistic star is given by,
%\sukanta{(Do we really want to use this signature as opposed to (-1,1,1,1)? Also, is this signature choice consistent with all the relevant equations in this paper?)}\bhaskar{(Yes, this signature choice is consistent with all the relevant equations in this paper. This signature is also used by Thorne and campolattaro \cite{Thorne1967} and Finn \cite{Finn1990}. This signature is chosen because we can verify our equations with Finn's paper.)}
\begin{equation}
ds^{2}=g_{\alpha \beta}dx^{\alpha}dx^{\beta}=e^{\nu(r)}dt^{2}-e^{\lambda(r)}dr^{2}-r^{2}d\theta^{2}-r^{2}\sin^{2}\theta d\phi^{2},
\end{equation}
where $\nu$ and $\lambda$ are two metric functions, and $\lambda$ can be expressed in terms of mass $m(r)$ inside a radius of r,
\begin{equation}
e^{\lambda(r)}=\left[1-\frac{2m(r)}{r}\right]^{-1} .
\end{equation}
Here we are interested in neutron stars which have a fluid core and a solid crust. We assume that in equilibrium configuration the contribution of shear stress due to the presence of solid crust vanishes~\cite{Finn1990,Penner2011}. In reality, this assumption is not necessarily correct. But since we are interested in small amplitude perturbation we can think that the background shear is almost negligible and its contribution is important only in the perturbed configuration. Therefore, the contribution of shear stress only enters through the perturbed stress energy tensor. The advantage of this assumption is that it makes our background problem very simple as we can now use the perfect fluid stress-energy tensor to model the background star:
\begin{equation}
T_{\alpha \beta}=(\rho+P)u_{\alpha}u_{\beta}-Pg_{\alpha \beta}\, ,
\end{equation}
where $u_{\alpha}$, $\rho(r)$ and $P(r)$ denote fluid 4-velocity, energy density and pressure, respectively, inside the star.
Solving Einstein equation for this equilibrium configuration we arrive at TOV equation,
\begin{equation}
 \frac{dP(r)}{dr}=-\frac{\left[\rho (r) + P(r) \right]\left[m(r) + 4\pi r^3 P(r) \right]}{r\left[r -2m(r)\right]}
  \label{tov1:eps}
\end{equation}
\begin{equation}
\frac{d\nu (r)}{dr} = - \frac{1}{\rho (r) + P(r)} \frac{dP (r)}{dr}
\label{tov2:eps}
\end{equation}
\begin{equation}
\frac{dm(r)}{dr}=4\pi r^{2}{\rho }(r).
    \label{tov3:eps}
\end{equation}
For a cold neutron star it is reasonable to assume that this fluid does not exchange heat with the surroundings. Therefore, one can take the EOS to be a zero-temperature barotrope: $P=P(\rho)$. Given the EOS of neutron stars, Eqs.~(\ref{tov1:eps}) and (\ref{tov3:eps}) can be solved to obtain their mass-radius relationship.

\subsection{Even parity perturbation}
\subsubsection{Fluid perturbation equations }
First we consider $l=2$, static, even-parity perturbations in the
Regge-Wheeler gauge. Since we are only interested in quadrupole deformation, $l=2$ case is considered from the beginning. Also we further simplify equation of motion by choosing spherical harmonics with $m=0$. Under these assumptions perturbed metric becomes,
\begin{equation}\label{Metric_Perturbed}
	h_{\alpha \beta}(r) = 
	\left(
	\begin{array}{cccc}
	H_0(r) e^\nu & 0 & 0 & 0 \\ 
	0 & H_2(r) e^\lambda & 0 & 0 \\ 
	0 & 0 & r^2 K(r) & 0 \\ 
	0 & 0 & 0 & r^2  \sin^2\theta K(r)
	\end{array} 
	\right)P_{2}(\cos\theta).
	\end{equation}
    The contravariant component of fluid displacement field takes the form
\begin{equation}
\xi^{r}=\frac{e^{-\lambda/2}}{r^{2}}WP_{2}(\cos\theta)\,,
\end{equation}
\begin{equation}
\xi^{\theta}=-\frac{V}{r^{2}}\partial_{\theta} P_{2}(\cos\theta)\,.
\end{equation}
For the case of perfect fluid all the off-diagonal components of perturbed stress-energy tensor vanish. The non-vanishing components of perturbed stress-energy tensor are
\begin{center}
$\delta T_{0}^{0}=\delta \rho$\,,
\end{center}
\begin{center}
$\delta T_{i}^{i}=-\delta P$\,.
\end{center}
For a barotrope we can assume the following form of perturbed pressure:
\begin{center}

$\delta P =\frac{dP}{d\rho} \delta \rho=c_{s}^{2} \delta \rho$\, ,
\end{center}
where $c_{s}^{2} $ is the speed of sound inside the star.

The set of equations which describes the fluid problem is given by
\begin{equation}
W^{'}=\frac{r^{2}e^{\lambda/2}}{2}(-K+H_{0})+\frac{3W}{r}+3V e^{\lambda/2}
\end{equation}
\begin{equation}
V^{'}=\frac{e^{\lambda/2}W}{r^{2}}+\frac{2V}{r}
\end{equation}
\begin{equation}
K^{'}=H_{0}\nu^{'}+H_{0}^{'}
\end{equation}
\begin{equation}\label{Fluid_Perturbation_01}
	\begin{split}
	&H_0^{''}+ H_0^{'} \bigg[\frac{2}{r} + e^{\lambda} \Big(\frac{2m(r)}{r^2} + 4 \pi r (P - \rho)\Big)\bigg] \\ 
	&+ H_0 \left[-\frac{6 e^{\lambda}}{r^2} + 4\pi e^{\lambda} \left(5 \rho + 9 P + \frac{\rho + P}{{c_s}^2}\right) - {\nu^\prime}^2\right] = 0,
	\end{split}
	\end{equation}
where the prime ($^{'}$) denotes derivative respect to $r$.	Basically, only a single differential equation, namely of  $H_{0}$ is sufficient to determine the tidal Love number of a fluid star. Rest of the three coupled differential equations are needed just to join the fluid core of the star with the solid crust.
\subsubsection{Elastic perturbation equations}
We assume our background star to be relaxed and unstrained. An elastic crust does not affect the equilibrium model. The contribution of elasticity comes only through the perturbed stress-energy tensor. Therefore, the perturbed stress energy tensor in solid medium becomes,
\begin{center}
$\delta T_{\alpha \beta}=\delta T_{\alpha \beta}^{fluid} +\delta \Pi_{\alpha \beta}$\, ,
\end{center}
where $\delta \Pi_{\alpha \beta}$ is anisotropic stress energy tensor. Detailed derivation of calculating this anisotropic stress energy tensor is given by Finn and Penner et al. \cite{Finn1990,Penner2011}. The non-vanishing components of $\delta \Pi_{\alpha \beta}$ are,
\begin{equation}
\delta \Pi_{r}^{r}= A Y_{lm}=\frac{2\mu}{3}\left[K-H_{2}+2re^{-\lambda/2}\left(\frac{W^{'}}{r^{3}}-\frac{3W}{r^{4}}\right)-\frac{l(l+1)V}{r^{2}}\right] Y_{lm} \, 
\label{pirr}
\end{equation}
\begin{equation}
\delta \Pi^{r}_{\mathcal{A}}=B Y_{lm,\mathcal{A}}=\mu\left[\frac{e^{-\frac{\lambda}{2}}}{r^{2}}W-r^{2}e^{-\lambda}\left(\frac{V^{'}}{r^{2}}-\frac{2V}{r^{3}}\right)\right]Y_{lm,\mathcal{A}}\, 
\label{pirt}
\end{equation}
\begin{equation}
\delta \Pi_{\mathcal{B}}^{\mathcal{A}}=2\mu V Y_{lm}\vert_{\mathcal{B}}^{\mathcal{A}} +\frac{\mu}{3}\left[H_{2}-K-2re^{-\lambda/2}\left(\frac{W^{'}}{r^{3}}-\frac{3W}{r^{4}}\right)-2\frac{l(l+1)V}{r^{2}}\right]\delta_{\mathcal{B}}^{\mathcal{A}} Y_{lm}\ ,
\end{equation}
where $\mathcal{A}$ and $\mathcal{B}$ run over the coordinates $\theta$ and $\phi$. The vertical ($\vert$) denotes covariant derivatives on two sphere.
Also we have defined two terms $A$ and $B$ which are related to radial and tangential component of the anisotropic shear stress tensor. Now
equation (\ref{pirr}) and (\ref{pirt}) can be used to integrate $W$ and $V$ (only $l=2$, $m=0$ case has been considered here):
\begin{equation}
W^{'}=\frac{r^{2}e^{\lambda/2}}{2}(\frac{3}{2 \mu}A-K+H_{0})+\frac{3W}{r}+(16\pi \mu r^{2}+3)Ve^{\lambda/2}
\label{W}
\end{equation}

\begin{equation}
V^{'}=\frac{e^{\lambda/2}W}{r^{2}}+\frac{2V}{r}-\frac{Be^{\lambda}}{\mu}
\label{V}
\end{equation}
In Regge-Wheeler gauge the perturbed number density takes the form \cite{Thorne1967}:
\begin{equation}
\frac{\Delta n}{n}=\left[-\frac{e^{-\lambda/2}}{r^{2}}W^{'}-\frac{6V}{r^{2}}+\frac{H_{2}}{2}+K\right]\, ,
\label{deltan}
\end{equation}
and the corresponding Lagrangian changes in density and pressure are:
\begin{equation}
\Delta \rho=(\rho +P)\frac{\Delta n}{n}
\label{deltar}
\end{equation}
\begin{equation}
\Delta P=c_{s}^{2}\Delta \rho .
\label{barotropic}
\end{equation}
The Lagrangian change in pressure is related to the Eulerian change as:
\begin{equation}
\Delta P=\delta P+\xi^{r}P^{'}=\delta P-\frac{(\rho +P)\nu^{'}}{2r^{2}}e^{-\lambda/2}W .
\label{Deltap}
\end{equation}
Combining (\ref{deltar}), (\ref{barotropic}) and (\ref{Deltap}) we get an expression for the perturbed Euler pressure:
\begin{equation}
\delta P=(P+\rho)c_{s}^{2}\left[-\frac{3}{4 \mu}A+\frac{3}{2}K-\frac{9V}{r^{2}}+\frac{e^{-\lambda/2}}{r^{3}}\left(-3+\frac{r\nu^{'}}{2c_{s}^{2}}\right)W\right].
\label{deltap}
\end{equation}
The $[rr]$ component of perturbed Einstein tensor gives another expression for $\delta P$:
\begin{align}
16\pi r^{2}e^{\lambda}(\delta P- A)&=4e^{\lambda}K-H_{0}[6e^{\lambda}-2+r^{2}(\nu^{'})^{2}]-r^{2}\nu^{'}H_{0}^{'}\nonumber\\
&-16\pi \mu V r^{2}(\nu^{'})^{2}+16\pi e^{\lambda} r B(2+r\nu^{'}).  
\label{rr}
\end{align}
By solving these two algebraic equations, (\ref{deltap}) and (\ref{rr}), we calculate $\delta P$ and $A$.
$[r \theta]$ component leads to equation of motion for $K$:
\begin{equation}
K^{'}=H_{0}\nu^{'}+H_{0}^{'}+\frac{16 \pi \mu (r\nu^{'}+2)V}{r}-16\pi B e^{\lambda}.
\label{K}
\end{equation}
Subtraction of $[\phi \phi]$ component from $[\theta \theta]$ component leads to:
\begin{equation}
H_{2}=H_{0}+32\pi \mu V .
\label{H2}
\end{equation}
We write the sum of $[\theta \theta]$ and $[\phi \phi]$ component in terms of $A$ and $B$
\begin{equation}
-\delta P=\frac{e^{-\lambda}}{16\pi r}(\nu^{'}+\lambda^{'})H_{0}-\frac{4 \mu V}{r^{2}} -\frac{B}{2r}(4+r\lambda^{'}+r \nu^{'}) -B^{'}+\frac{A}{2}\, .
\end{equation}
We can use this equation to integrate $B$:
\begin{equation}
B^{'}=\frac{e^{-\lambda}}{16\pi r}(\nu^{'}+\lambda^{'})H_{0} -\frac{B}{2r}(4+r\lambda^{'}+r \nu^{'})-\frac{4 \mu V}{r^{2}}+\delta P +\frac{A}{2} \,.
\label{B}
\end{equation}
If we take the trace of perturbed Einstein equation we arrive at a second order equation for $H_{0}$:
\begin{multline}
-r^{2}H_0^{''}+\left(\frac{1}{2}r\lambda^{'}-r\nu^{'}-2\right)rH_{0}^{'}+r^{2}\nu^{'}K^{'}-\frac{1}{2}r^{2}\nu^{'}H_{2}^{'}+6e^{\lambda}H_{0}\\
+[2(e^{\lambda}-1)-r(\lambda^{'}+3\nu^{'})]H_{2}=-8\pi r^{2}e^{\lambda}(3\delta P+\delta \rho)\,.
\end{multline}
After plugging (\ref{K}) and (\ref{H2}) in the above equation we get
\begin{multline}
-r^{2}H_0^{''}+\left(\frac{1}{2}r(\lambda^{'}-\nu^{'})-2\right)rH_{0}^{'}+[6e^{\lambda}+2(e^{\lambda}-1) 
-r(\lambda^{'}+3\nu^{'})+r^{2}(\nu^{'})^{2}]H_{0}  \\
=8\pi\{ -r^{2}e^{\lambda}(3\delta P+\delta \rho) 
+8\mu\left[1-e^{\lambda}+r\left(\nu^{'}+\frac{1}{2}\lambda^{'}\right)-\dfrac{1}{4}(r\nu^{'})^{2}\right]V\\
+2r^{2}\nu^{'}(\mu V)^{'}+2r^{2}\nu^{'}B e^{\lambda}\} \,.
\label{H}
\end{multline}
The five differential equations \eqref{W}, \eqref{V}, \eqref{K}, \eqref{B} and \eqref{H} given above together with two algebraic equations \eqref{deltap} and \eqref{rr} form a complete set of equations which describe the evolution of perturbed quantities in the elastic medium of the star.

\subsubsection{Boundary condition at center and stellar surface}

At the center of the star, all the perturbed quantities must be regular. For our study we take core of the star to be fluid, for which the boundary conditions were analyzed by Thorne and Campolattaro \cite{Thorne1967}. Here we just summarize their result. All the perturbed quantities are expanded in Taylor series about $r=0$ as:
\begin{eqnarray}
H_{0}&=&r^{l}[H_{0}^{(0)}+H_{0}^{(2)}r^{2}+...]\, ,\label{eq:H00}\\
K&=&r^{l}[K^{(0)}+K^{(2)}r^{2}+...]\, ,\label{eq:K0}\\
W&=&r^{l+1}[W^{(0)}+W^{(2)}r^{2}+...]\, ,\label{eq:W0}\\
V&=&r^{l}[V^{(0)}+V^{(2)}r^{2}+...]\, \label{eq:V0}.
\end{eqnarray}
Using these expansions in equation (\ref{deltan}) we get $W^{(0)}=-lV^{(0)}$. Next, by combining equations (\ref{rr}) and (\ref{B}) for $\mu =0$ we obtain  :
\begin{equation}
4e^{\lambda}K-H_{0}[6e^{\lambda}-2+r^{2}(\nu^{'})^{2}-r(\nu^{'}+\lambda^{'})]-r^2\nu^{'}H_0^{'}=0\, .
\end{equation}
It is straightforward to show that expansion of this equation about $r=0$ leads to $K^{(0)}=H^{(0)}_{0}$. Therefore, out of  four constants appearing in equations (\ref{eq:H00})-(\ref{eq:V0}), only two are independent. These two are fixed by the demand that the Lagrangian perturbation of pressure vanishes at the surface of the star.

\subsubsection{Interface condition}

We have derived the perturbation equations in the solid crust region. Now we need to find proper interface conditions to join them with the fluid perturbation equations in the core. The interface conditions are obtained from the equations of motion of fluid variables and Einstein Field equation (please see \cite{Finn1990} for the derivation). The continuity of intrinsic curvature demands that $H_{0}$, $K$, $W$ must be continuous at the interface:
\begin{eqnarray}
&&[H_{0}]_{r_i} = 0\,, \\
&&[K]_{r_i}  = 0\,, \\
&&[W]_{r_i} = 0\,.
\end{eqnarray}
Again, continuity of extrinsic curvature imposes two additional boundary conditions:
\begin{equation}
[\Delta P -A]_{r_{i}}=0\, ,
\label{ex1}
\end{equation}
\begin{equation}
[B]_{r_{i}}=0 .
\end{equation}
Since, $W$ is continuous across the interface equation (\ref{ex1}) reduces to :
\begin{equation}
[\delta P -A]_{r_{i}}=0\, .
\end{equation}
By noting that  $A=0$ in the fluid core we  obtain the value of radial stress at the interface as :
\begin{equation}
A_{i}=\delta P_{i}-\delta P_{f} \, ,
\end{equation}
where $\delta P_{f}$ and $\delta P_{i}$  are the Eulerian perturbations of pressure at the base of fluid core and at the interface, respectively. Using equations (\ref{deltap}) and (\ref{B}) we get the expression of $\delta P_{f}$ and $\delta P_{i}$ :
\begin{equation}
\delta P_{f}=\frac{1}{2}(\rho+P)H_{0f} \, ,
\end{equation}
\begin{equation}
\delta P_{i}=(P+\rho)c_{s}^{2}\left[-\frac{3}{4 \mu } A_{i}+\frac{3}{2}K_{i}-\frac{9V_{i}}{r^{2}}+\frac{e^{-\lambda/2}}{r^{3}}\left(-3+\frac{r\nu^{'}}{2c_{s}^{2}}\right)W_{i}\right]\, .
\end{equation}

\subsubsection{Calculation of electric tidal Love number}
 Our focus here is to calculate the electric love number of neutron stars consisting of a fluid core and an elastic crust. We first integrate the
 fluid perturbation equations starting from the center of the star to the 
 core-crust junction, using the specified boundary conditions at the center. Next,
 we integrate the elastic perturbation equations from this junction to the surface.
 The starting point of the later integration is obtained by imposing the interface 
 conditions at the core-crust junction. 
 Now, in order to calculate the tidal Love number we have to match this internal solution with the external solution at the surface of the star. We suggest the reader to see Refs. \cite{Hinderer2008,Binnington2009,Damour2009}, where extensive details about the calculation of tidal Love number can be found. The value of tidal Love number can be computed in terms of $y$ and compactness parameter $C=\frac{M}{R}$ as :
\begin{equation}
\label{expr_k2}
\begin{split}
k_2 &= \frac{8}{5}(1-2C)^2C^5\big[2C(y-1)-y+2\big]\bigg[2C(4(y+1)C^4 \\
&+ (6y-4)C^3+(26-22y)C^2+3(5y-8)C-3y+6) \\
&-3(1-2C)^2(2C(y-1)-y+2)\log(\frac{1}{1-2C})\bigg]^{-1}\, ,
\end{split}
\end{equation}
where $y$ depends on the value of $H_{0}$ and its derivative at the surface:
\begin{equation}
    y=\left.\frac{rH_{0}^{'}}{H_{0}}\right\vert_{R}\,.\label{eq:y}
\end{equation}

\subsection{Odd parity perturbation:}
%We next study the odd parity perturbations.
\subsubsection{Fluid perturbation equations:}
Magnetic tidal Love number were  computed together by Binnington and Poisson~\cite{Binnington2009} (BP) and Damour and Nagar~\cite{Damour2009} (DN) back in 2009. In their calculation, BP assumed that tidal field varies slowly over the time, therefore it never throws the body out of hydrostatic equilibrium. Based on this assumption they derived all the perturbation equations using a static-fluid ansatz and from there they calculated magnetic tidal Love number. On the other hand, instead of re-deriving the perturbation equations, DN took the  Cunningham, Price and
Moncrief master function~\cite{Cunningham1978} governing odd parity perturbation of Schwarszchild space-time and used a stationary perfect-fluid ansatz for stress-energy tensor. In 2015 Landry and Poisson~\cite{Landry2015} (LP) revisited BP's calculation by taking an irrotational state of the fluid which permits internal motions of fluid inside the body. They find that magnetic tidal Love number for this irrotational state are different from the magnetic Love number associated with static fluid. LP also found that there results agree with DN's result since irrotational condition is automatically imposed by the stationary master function chosen by DN. This state of affair is recently re-examined by Pani et al.~\cite{Pani2018}. 
 In our work, we allow internal motion of fluid since it is the more realistic configuration to describe the fluid than the hydrostatic equilibrium scenario.

 We consider here  magnetic type perturbation for $l=2, m=0$ in the Regge-Wheeler gauge by a time dependent tidal field. However we assume tidal field varies very slowly over the time, therefore, we neglect all the time derivative appears in our field equations. But this slowly varying tidal field does have impact on the internal motion of the fluid which establishes irrotational state of it. Under the above mentioned assumption the perturbed metric becomes:
\begin{equation}\label{odd_Metric_Perturbed}
	h_{ab}(r) = 
	\left(
	\begin{array}{cccc}
	0 & 0 & 0 & h_{0}(r,t) \\ 
	0 & 0 & 0 & h_{1}(r,t) \\ 
	0 & 0 & 0 & 0 \\ 
	h_{0}(r,t) & h_{1}(r,t)& 0 & 0
	\end{array} 
	\right)\sin\theta\partial_{\theta}P_{2}(\cos\theta).
	\end{equation}
	The contravariant fluid displacement vector has the following form:
	\begin{equation}
	    \xi_{r}=\xi_{\theta}=0, \hspace{10ex} \xi_{\phi}=U(r,t)\sin{\theta} \partial_{\theta}P_{2}(\cos{\theta})\, ,
	\end{equation}
    where $U(r,t)$ is the fluid displacement function for odd parity perturbation. Perturbed four-velocities corresponds to this first order in displacement, are
    \begin{equation}
	    v_{r}=v_{\theta}=0, \hspace{10ex} v_{\phi}=e^{-\nu/2}U_{,t}\sin{\theta} \partial_{\theta}P_{2}(\cos{\theta})\, ,
	\end{equation}
	Since density and pressure are scalar they do not change under odd parity perturbation. However fluid four velocity will be shifted to $u^{\mu}$ to $u^{\mu}+\delta u^{\mu}$. In first order perturbation we note the following relation, $\delta u_{\mu} =g_{\mu \nu} \delta u^{\nu}+h_{\mu \nu}u^{\nu}$ to compute the components of $\delta u_{\mu}$,
	\begin{equation}
	    \delta u_{r}=v_r, \hspace{5ex}
	    \delta u_{\theta}=v_{\theta},
	    \hspace{5ex}
	    \delta u_{\phi}=v_{\phi}+h_{t\phi}u^t\, ,
	\end{equation}
	where, $v_{\mu}=g_{\mu \nu}\delta u^{\nu}$. Now, irrotational state of fluid implies $\delta u_r=0=\delta u_{\mathcal{A}}$~\cite{Landry2015}, where $\mathcal{A}$ runs over the coordinate $\theta$ and $\phi$. Consequently, the form of perturbed stress-energy tensor can be written as,
	\begin{equation}
	    \delta T_{\nu}^{\mu}=(\rho + P)(u^{\mu}\delta u_{\nu}+\delta u^{\mu}u_{\nu})-P\delta_{\nu}^{\mu}
	\end{equation}
	Therefore, for the irrotational case,
	the $[t\phi]$ component of Einstein equation give us, 
	\begin{equation}
	    h_0^{''}-\frac{\lambda^{'}+\nu^{'}}{2}h_0^{'}-\left[\frac{4e^{\lambda}}{r^{2}}+\frac{2}{r^{2}}-\frac{\lambda^{'}+\nu^{'}}{r}\right]h_0=0
	        \label{h0}\,,
	\end{equation}
	        For the static case, $v_{\phi}=0$ gives $\delta u_{\phi}=h_{t\phi}u^t$. In that case, the $[t\phi]$ component of Einstein equation give us,
	        	\begin{equation}
	    h_0^{''}-\frac{\lambda^{'}+\nu^{'}}{2}h_0^{'}-\left[\frac{4e^{\lambda}}{r^{2}}+\frac{2}{r^{2}}-\frac{\lambda^{'}+\nu^{'}}{r}\right]h_0=0
	        \label{h0_static}\,,
	\end{equation}
	Notice that assumption of irrotational state of fluid changes the sign of $\frac{\lambda^{'}+\nu^{'}}{r}$ in the term which is proprtinal to $h_0$.
	
	    Since, we will be working with irrotational state of fluid, We need to solve (\ref{h0}) with the regular boundary condition at the center:
\begin{equation}	        h_{0}=h_{0}^{(0)}r^{3}+\mathcal{O}(r^{5})\,,
\end{equation}
where $h_{0}^{(0)}$ is an arbitrary constant.

\subsubsection{Elastic perturbation equations:}
	        
In this case the non-vanishing components of the anisotropic stress energy tensor are:
\begin{eqnarray}
    \delta \Pi_{\phi}^{t}&=&-\mu e^{-\nu} h_{0}\sin{\theta}\partial_{\theta}P_{2}(\cos{\theta})\\
    \delta \Pi_{\phi}^{r}&=&\mu e^{-\lambda}\left[U^{'}-\frac{2U}{r}+h_{1}\right]\sin{\theta}\partial_{\theta}P_{2}(\cos{\theta})\\
    \delta \Pi_{\phi}^{\theta}&=&\frac{3\mu}{r^{2}}U\sin^{3}{\theta}\,.
  \end{eqnarray}
After including these anisotropic stress energy-tensor in the perturbed Einstein equations the $[t\phi]$ gives us:
\begin{equation}
h_0^{''}-\frac{\lambda^{'}+\nu^{'}}{2}h_0^{'}+\left[\frac{\lambda^{'}+\nu^{'}}{r}-\frac{4e^{\lambda}}{r^{2}}-\frac{2}{r^{2}}+16\pi \mu e^{\lambda}\right]h_0=0
	        \label{h0c}
\end{equation}
  We  solve equation (\ref{h0}) from the center to core-crust junction and equation (\ref{h0c}) from there to the stellar surface where at the core-crust junction $h_0$ is continuous.

 \subsubsection{Magnetic Love number calculation:}
	        
	Asymptotic behavior of  equation \eqref{h0} at large distances is given by,
	\begin{equation}
	    \left(1-\frac{2M}{r}\right)h_0^{''}+\left[-\frac{6}{r^2}+\frac{4M}{r^3}\right]h_0=0
	\end{equation}
%\begin{widetext}
By matching its asymptotic solution with $g_{t\phi}$ component of metric in asymptotically mass-centered Cartesian coordinate (ACMC)\cite{Thorne1980,Cardoso2017}, we obtain the expression of magnetic Love number
\begin{equation}
\label{mag_expr_k2}
\begin{split}
j_2 = \frac{8C^5}{5}\frac{2C (y-2)-y+3}{2C \left[2C^3 (y+1) + 2C^2 y + 3C (y-1)-3 y+9\right] + 3[2C (y-2)-y+3] \log (1-2C)}, \,
\end{split}
\end{equation}
where $y=\frac{r h_{0}^{'}}{h_0}$ is evaluated at the surface of the star and $C$ is the compactness.

\section{Equation of state}
\label{EOS}

\begin{figure}[ht]
\begin{tabular}{cc}
\includegraphics[width=0.48\textwidth]{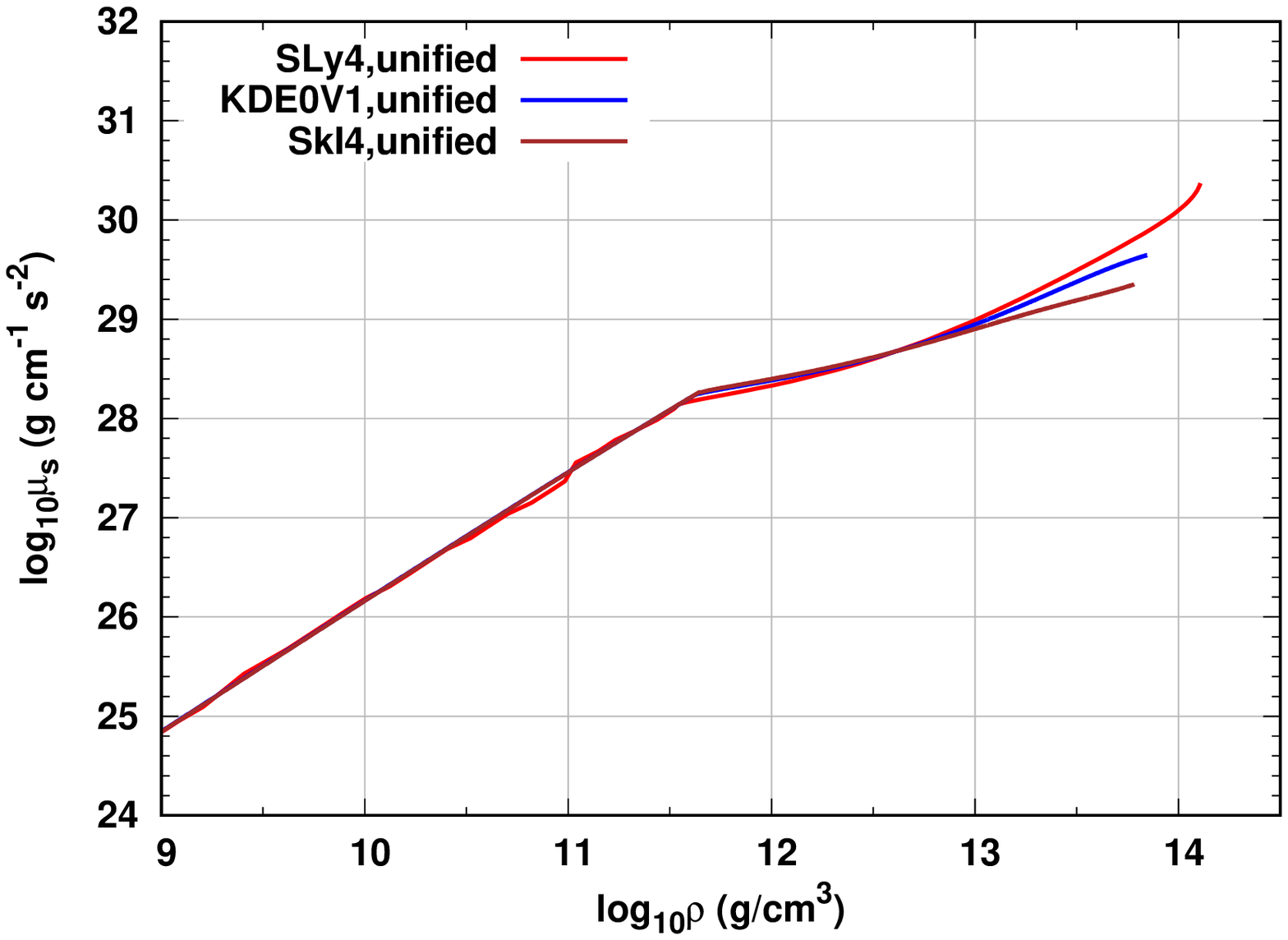}&
\includegraphics[width=0.48\textwidth]{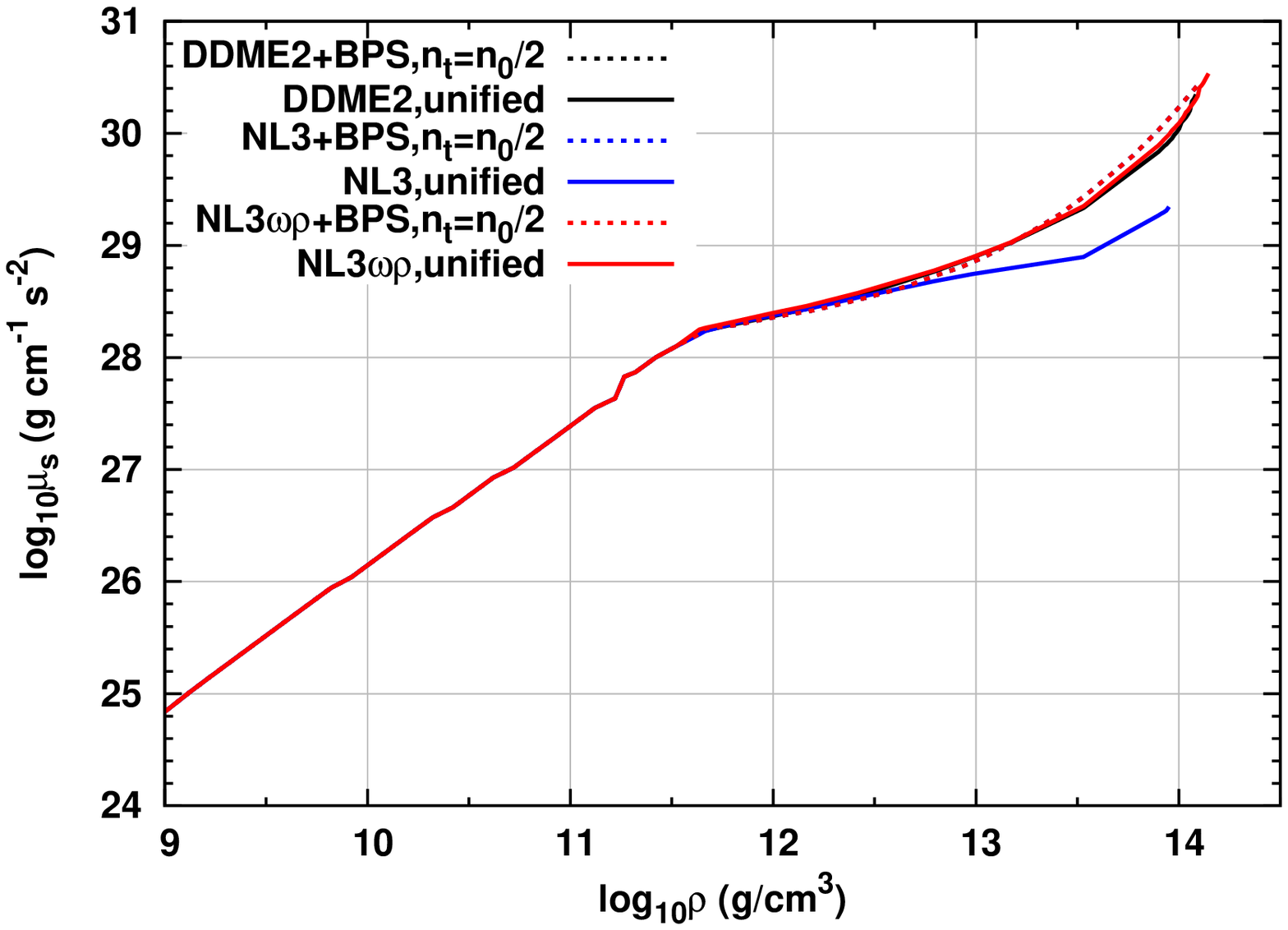}\\
\end{tabular}
\caption{Shear modulus vs $\rho$ for non-relativistic (left panel) and relativistic EOSs (right panel).}
\label{fig:shear}
\end{figure}

It has been shown \cite{Fortin2016} that for an unambiguous calculation of NS properties (especially radius and crust thickness) it is necessary to adopt unified equations of state (EOS), where the EOS of crust and core are obtained within the same many-body theory. As the crust thickness plays the key role in the present study we employ only unified EOSs here. We consider six unified EOSs: SLy4 \cite{Douchin:2001sv},  KDE0V1 \cite{Agrawal:2005ix}, SkI4 \cite{Reinhard:1995zz}, NL3 \cite{Lalazissis:1996rd,Grill:2014aea}, NL3$\omega\rho$ \cite{Horowitz:2002mb,Grill:2014aea} and DDME2 \cite{Lalazissis:2005de,Grill:2014aea}. The first three are based on non-relativistic Skyrme interactions and are
obtained from the CompOSE database~\cite{compose,Gulminelli:2015csa}. The
other three are derived from the relativistic mean-field (RMF) model. The RMF EOSs are not fully unified as the outer crust is not calculated within the same theory but is taken from Ref. \cite{Baym:1971pw}.
Since the most part of the outer crust is determined from the experimentally measured nuclear masses, the choice of it does not significantly affect the observables.
The important properties of these EOSs are shown in Table \ref{tab:eos}. All of them are consistent with the observed maximum mass ($2.01\pm0.04M_\odot$) of neutron stars \cite{Antoniadis:2013pzd}. The shear moduli for all the EOSs are
plotted in Fig. \ref{fig:shear}. They are 
calculated using the following expression \cite{1991ApJ...375..679S,2016PhRvC..94b5801N}:
\begin{equation}
\mu = 0.1194\frac{n_i(Ze)^2}{a},
\end{equation}
where $a=[3/(4\pi n_i)]^{1/3}$, $n_i$ is the density of ions and $Z$ is the atomic number of the nucleus present.

 \begin{table}[ht]
\centering
\caption{Properties of unified EOSs are tabulated here. Specifically, $n_0$ is the saturation density, $n_t$ is the crust-core transition density, $K$ is the incompressibility, $J$ and $L$ are symmetry energy and its slope at saturation density, respectively.}
\label{tab:eos}
\begin{tabular}{cccccccc}
\hline 
EOS & $n_0$(fm$^{-3}$) & $K$(MeV) & $J$(MeV) & $L$(MeV) &  $n_t$(fm$^{-3}$) & $M_{\rm max}/M_\odot$ & $\mu_{n=n_t} (\rm g  cm^{-1} s^{-2}) $\\
 \hline\hline 
 SLy4 \cite{Douchin:2001sv}  & 0.159  & 230.0 & 32.0 & 46.0 &  0.0800 & 2.05 & 2.34 $\times 10^{30}$\\
% BSK21\cite{Goriely:2010bm} & 0.1582 & 245.8 & 30.0  & 46.6  & 0.0809 & 2.27\\
 KDE0V1 \cite{Agrawal:2005ix}& 0.165  & 227.5& 34.6 & 54.7 & 0.0480 & 1.97 & $4.43 \times 10^{29}$\\
  SkI4 \cite{Reinhard:1995zz}&  0.160 & 248.0  & 29.5 & 60.4  & 0.0359 & 2.18 & $2.24 \times 10^{29}$\\
 \hline 
NL3 \cite{Lalazissis:1996rd,Grill:2014aea} & 0.148 & 270.7 & 37.3&  118.3  & 0.0548 & 2.77 & $2.20 \times 10^{29}$ \\
 NL3$\omega\rho$ \cite{Horowitz:2002mb,Grill:2014aea} & 0.148 & 272.0 & 31.7 &55.3 &  0.0835 & 2.75 & $3.42 \times 10^{30}$\\
DDME2 \cite{Lalazissis:2005de,Grill:2014aea}& 0.152 & 250.9 & 32.3 & 51.2 &  0.0735 & 2.48 & $2.21 \times 10^{30}$\\
 \hline 
 NL3 matched & 0.148 & 270.7 & 37.3&  118.3  & 0.0740 & 2.77 & $2.73 \times 10^{30}$\\
 NL3$\omega\rho$ matched & 0.148 & 272.0 & 31.7 &55.3 &  0.0740 & 2.75 & $2.73 \times 10^{30}$\\
DDME2 matched& 0.152 & 250.9 & 32.3 & 51.2 &  0.0760 & 2.48 & $2.89 \times 10^{30}$\\
\hline
 \end{tabular}
 \end{table}

\section{Results}
\label{sec:results}

\begin{figure}[ht]
\begin{tabular}{cc}
\includegraphics[width=0.48\textwidth]{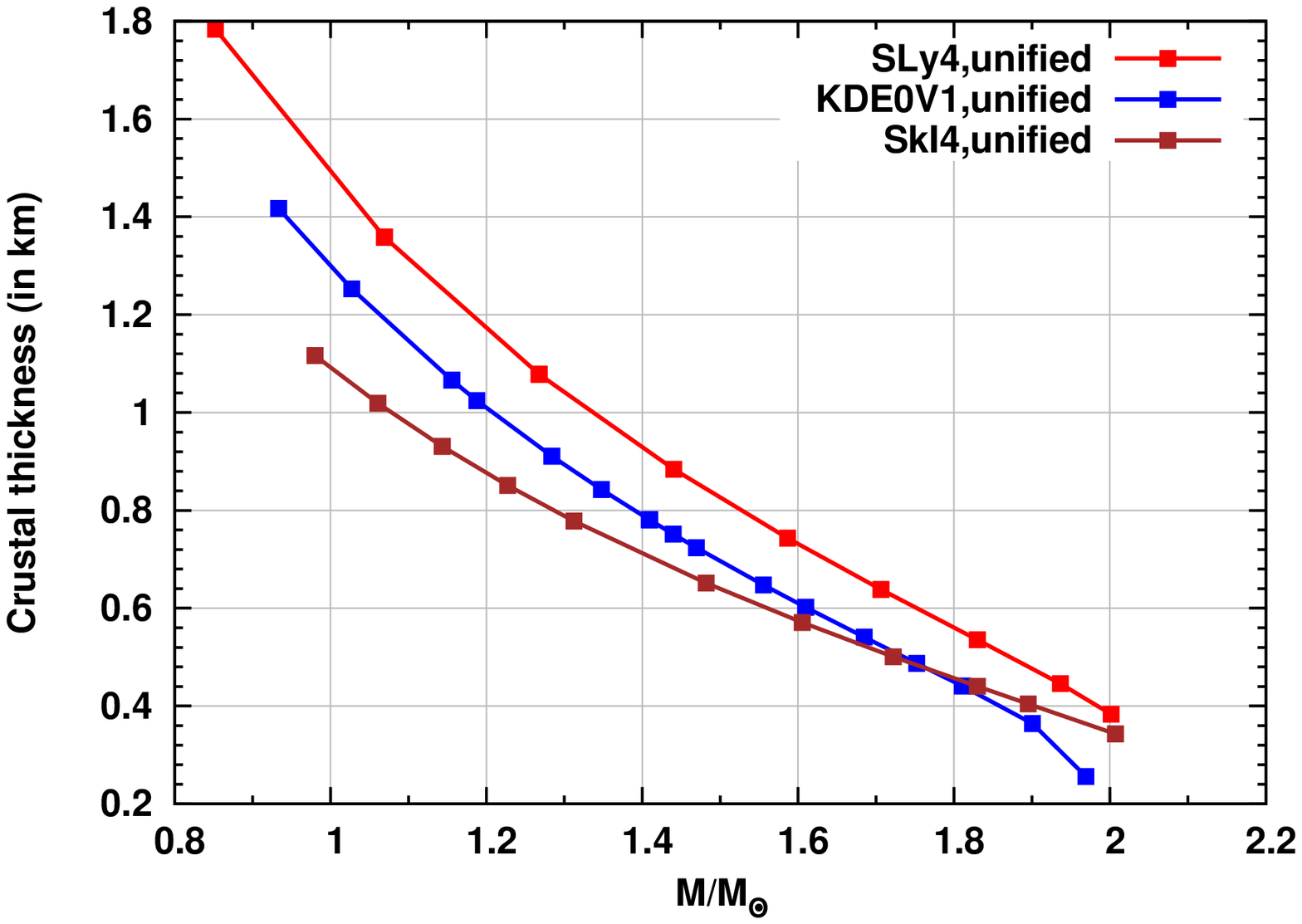}&
\includegraphics[width=0.48\textwidth]{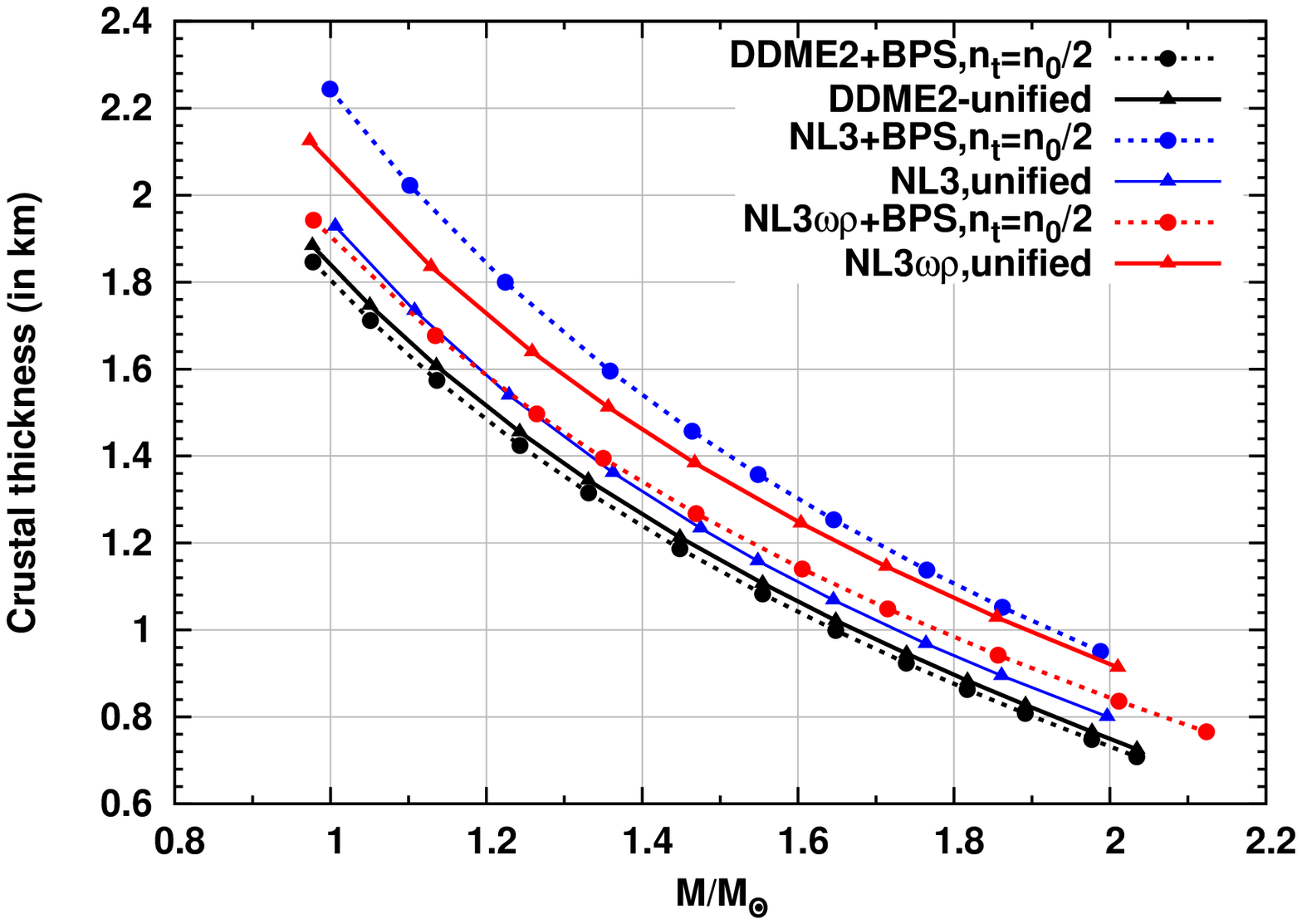}\\

\end{tabular}
\caption{Crustal thickness vs mass for non-relativistic (left panel) and relativistic EOSs (right panel).}
\label{fig:crustal thickness}
\end{figure}

In this section we present our numerical findings for a set of neutron star EOSs that were discussed in Sec.~\ref{EOS}. First, for each of these EOSs we generated a set of equilibrium stellar configurations within the mass range of $1M_{\odot}$ to $2M_{\odot}$ by solving the TOV equations. In Fig. \ref{fig:crustal thickness}, crustal thickness is plotted w.r.t mass for each considered EOSs. Then for each star we integrate all the perturbed variables and calculate the corresponding Love number.  For our numerical calculations, we use dimensionless variables in all the necessary differential equations, which are presented in the Appendix~\ref{appendix}. All the differential equations are solved using fourth order Runge-Kutta method.

\subsection{Even parity perturbations:}

\subsubsection{Non relativistic EOS}

\begin{figure}[ht]

\includegraphics[width=0.45\textwidth]{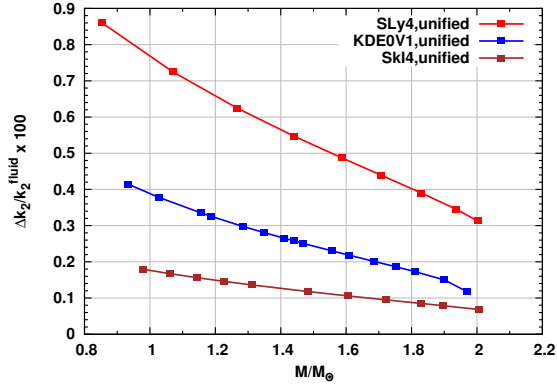}\\

\caption{Percentage change in $k_{2}$ vs mass for non-relativistic unified EOSs.}
\label{fig:Nonrel_electric_delk}
\end{figure}

In Fig. \ref{fig:Nonrel_electric_delk}, we plot the change in $k_2$ (in $\%$) due to the inclusion of the crust as a function of mass for three non-relativistic EOSs. 
The fractional change in $k_{2}$ is defined as $\Delta k_{2}/k_{2}^{\rm fluid}$, where $\Delta k_{2}=k_{2}^{\rm fluid}-k_{2}^{\rm crust}$; $k_{2}^{\rm fluid}$ and $k_{2}^{\rm crust}$ are, respectively, the tidal Love numbers of a purely fluid star and a star with an elastic crust. Since, the elastic crust would resist deformation,
it is expected that $k_2^{\rm crust}<k_2^{\rm core}$, resulting in 
$\Delta k_2>0$. This is indeed the case as can be seen from Fig. \ref{fig:Nonrel_electric_delk}.
It is also observed that as the thickness of the crust increases, the change in Love number increases (for comparison, see Fig. \ref{fig:crustal thickness}).
% That means the effect will be higher as the star becomes more elastic.
The change in $k_{2}$ is about $~0.1-0.4\%$ for KDE0V1 EOS, $~0.3-0.9\%$ for SLy4 EOS and $~0.1-0.2\%$ for SkI4 EOS. For a given mass, the increasing order of crustal thickness among these three non-relativistic unified EOSs is: SLy$4>$ KDE0V1 $>$ SkI4. A similar trend is seen 
%sort of increasing order is seen in 
for the  change in $k_{2}$ in Fig.~\ref{fig:Nonrel_electric_delk}, which points to the fact that stars with bigger crusts would have lesser deformation, as expected.

\subsubsection{Relativistic EOS}

\begin{figure}[ht]

\includegraphics[width=0.5\textwidth]{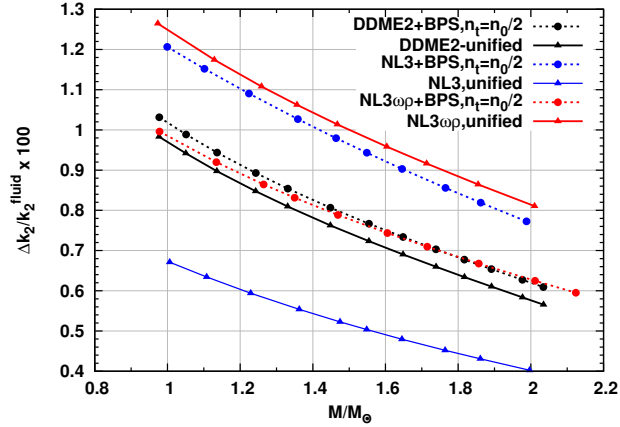}\\

\caption{Percentage change in $k_{2}$ vs mass for relativistic unified EOSs.}
\label{fig:electric_delk}
\end{figure}

%\rn{}
Change in $k_2$ as a function of mass is plotted in Fig.~\ref{fig:electric_delk} for three unified RMF EOSs. To investigate the importance of unified EOS, we also include EOSs obtained by matching a crust EOS with the core EOS in a thermodynamically consistent way~\cite{Fortin2016}.
% We use both unified and thermodynamically consistent matched EOSs.
% For thermodynamically matched EOSs, the RMF EOSs of core is
All three RMF EOSs of core are
matched to the BPS+BBP~\cite{Baym:1971pw, Baym:1971ax} EOS of the crust at $n_t=n_0/2$,
%{} 
where $n_0$ is the saturation density of the core EOS (see table \ref{tab:eos} for values). From Fig. \ref{fig:crustal thickness} we see that the crust is bigger for the matched EOS than that of unified EOS for NL3. Whereas, for DDME2 and NL3$\omega\rho$ the scenario is opposite. The reason is for unified NL3, $n_t=0.0548$ fm$^{-3}$ and for the matched case $n_t=n_0/2=0.074$ fm$^{-3}$. As the transition from crust to core happens later in the matched EOS, the crust is bigger in size for matched NL3. In contrast,  the transition happens earlier for matched EOS in case of DDME2 and NL3$\omega\rho$, which leads to a reduction in the crustal thickness. 

For non-relativistic EOSs we have already seen that larger crustal thickness gives larger deviation from the fluid value of $k_2$.  A similar behavior seen to be in play here. However, the magnitude of the shear modulus also plays a role in the change of $k_2$. Models with higher magnitude of shear modulus will be less deformed, which will correspond to a larger change in $k_2$. A careful inspection of Figs.~\ref{fig:shear} and \ref{fig:electric_delk} supports this finding. Depending on the magnitude of crustal thickness and shear modulus, one of the effects dominates over the other. For example, unified DDME2 has larger crustal thickness than matched DDME2, but the shear modulus of the latter is higher than that of the first. However, we find that the change in $k_2$ is higher for matched DDME2. Since the difference in crustal thickness between these two EOSs is very small, the change in $k_2$ is mainly caused
by the magnitude of the shear modulus in this case. For reference, we have included the values of the shear
modulus at the bottom of the crust in the last column of Table~\ref{tab:eos}. Overall,
the change in $k_2$ due to the presence of a solid crust is between $\sim 0.4-1.3\%$ for all the RMF EOSs considered here.
It is also noted that $\Delta k_2$ for
matched EOS can considerably differ from that of unified EOS. Among the three
RMF EOSs studied here we found that the difference is highest for the NL3 EOS and can be as large as  $\sim 90\%$. This emphasizes the necessity of the use
of unified EOSs in such calculations.
%therefore, effect of the magnitude of shear modulus dominates over crustal thickness.
  
\begin{figure}[ht]

\includegraphics[width=0.5\textwidth]{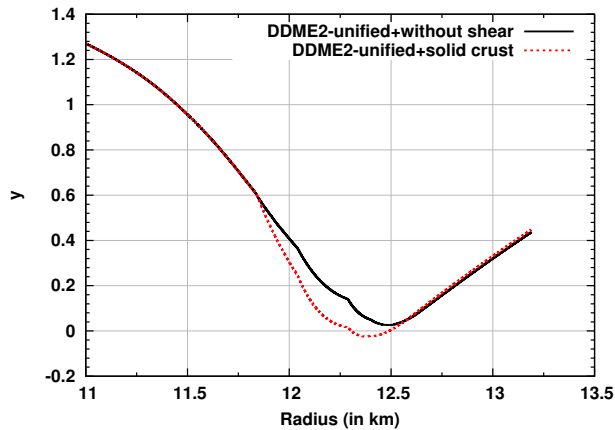}\\

\caption{$y$ is plotted as a function of radius for a 1.33$M_{\odot}$ neutron star using unified DDME2 EOS: solid black curve describes the profile of $y(r)$ for perfect fluid star and dotted red curve for a star whose crust is solid.  }
\label{fig:y}
\end{figure}

The above results show that even with realistic EOSs and realistic crustal models, shear modulus of solid crust has a small effect on the electric tidal Love number. The reason why the effect of shear modulus on $k_2$ is negligible can be understood from Fig.~\ref{fig:y}, where the profile of $y$ is plotted as a function of radius in the presence and absence of shear using unified DDME2 EOS. The dotted (red) and solid (black)  curves represent the cases of with and without shear, respectively. It is seen that the values of $y$ mainly differ from the fluid case in the inner crust region, while in the outer crust region difference from the fluid case is negligible. This happens because the magnitude of shear modulus is much higher in the inner crust and as a result the inner crust has greater response towards the tidal field than the outer crust. However, as the value of $y$ at the surface only enters into the calculation of $k_2$ (see eq. \ref{eq:y}),  we do not observe any significant change in it.

\subsection{Numerical results for odd parity perturbations:}
\subsubsection{Non-Relativistic EOS}

\begin{figure}[ht]
    \centering
    \includegraphics[width=0.5\textwidth]{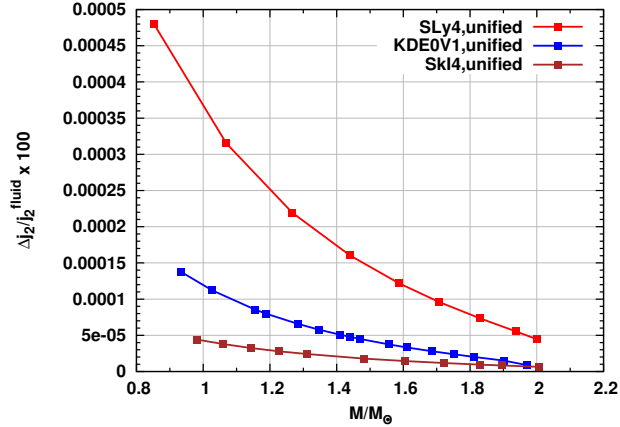}
    \caption{Percentage change in $j_{2}$ vs mass for non-relativistic unified EOSs.}
    \label{fig:mag_delk_nonrel}
\end{figure}

In Fig. \ref{fig:mag_delk_nonrel}, we plot the percentage change in magnetic Love number w.r.t mass for three unified non-relativistic EOSs. The change in magnetic Love number is  denoted as $\Delta j_2/j_{2}^{\rm fluid}$, where $\Delta j_2=\mid j_{2}^{\rm fluid}-j_{2}^{\rm crust} \mid$; $j_{2}^{\rm fluid}$ and $j_{2}^{\rm crust}$ are respectively magnetic Love number of a perfectly fluid star and a star with elastic crust. The value of magnetic Love number is itself negative and that's why the absolute values of difference have been taken.
%We observe different changes in $\Delta j_{2}$ for different EOSs. This happens mainly due to different values of shear modulus (see, Fig. \ref{fig:shear}) of different EOSs. If the magnitude of shear modulus is greater it's response towards the tidal field will be greater also i.e. larger change in $j_2$ w.r.t the fluid case, which is exactly seen here. 
Similar to the electric Love number, the magnetic
love number is also found to be affected by both the values of crustal thickness and the shear modulus. The change in $j_{2}$ varies between $\sim 0.00005-0.0005\%$ for considered non-relativistic EOSs.
This suggests deviation in magnetic Love number due to solid crust is negligible. %For magnetic Love number effect of shear modulus dominates over the effect of crustal thickness. This becomes more clear when we discuss the change in $j_2$ for relativistic EOSs, next.

\subsubsection{Relativistic EOS}
\begin{figure}[ht]

\includegraphics[width=0.5\textwidth]{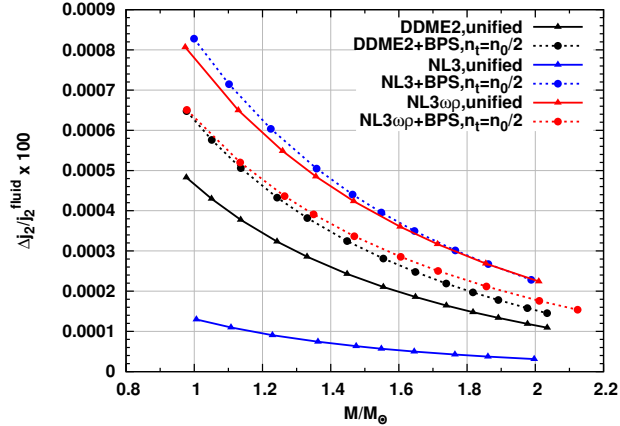}\\

\caption{Percentage of change in $j_{2}$ vs mass for relativistic unified EOSs.}
\label{fig:mag_delk}
\end{figure}
For RMF EOSs we observe similar type of changes in $j_2$.
From Fig. \ref{fig:shear} we see that the value of shear modulus is similar for all the matched EOSs but the crustal thickness is different (see, Fig \ref{fig:crustal thickness}) with values in the order
: NL3$>$NL$3\omega \rho>$DDME2. We see similar order of change in $j_2$ for matched EOSs. On the other hand, the change in $j_2$ is dominated by the magnitude of shear modulus for the unified EOSs as both the shear modulus (see, Fig. \ref{fig:shear}) and the change in $j_2$ have same order: NL$3\omega \rho>$DDME2$>$NL3. \
%We find changes of magnetic Love number in this order for the unified EOSs.
However, similar to non-relativistic EOSs, here also the changes in magnetic Love number are in between $\sim 0.0001-0.0009\%$ and hence, are practically negligible.

\subsection{Summary of the results:}

Here we briefly summarize the key findings of our numerical studies.
%\pc{(Is 'simulation' a proper choice of word here?)}. 
We have analyzed the effect of elastic crust on both electric and magnetic tidal Love numbers for a set of realistic equations of state and realistic models of shear modulus. In our study, we used 3 non-relativistic and 3 relativistic EOSs. Non-relativistic EOSs are based on Skyrme interactions and relativistic EOSs are constructed using relativistic mean field theory. We also used matched RMF EOS where core RMF EOSs are matched to BPS EOS of the crust at half of saturation density of the core EOS.

\begin{itemize}
\item Effect on electric Love number: Percentage change in $k_{2}$ is higher for relativistic EOSs than the non-relativistic ones. The reason is the RMF EOSs have larger crustal thickness. We observed that the EOSs with larger crustal thickness have a larger change in $k_{2}$ w.r.t the fluid case. Also we observe larger shear modulus corresponds to larger change in $k_2$. So we conclude that the crustal thickness and the magnitude of the shear modulus both have an effect on the electric tidal Love number to varying degrees depending on the EOS.

\item Effect on magnetic Love number: Similar to electric Love number both the crustal thickness and magnitude of shear modulus have effect on the magnetic Love number. Note, however, that the magnetic Love number is much smaller than the electric one for all of these EOSs and therefore it is highly unlikely that their imprints will be observed in BNS waveforms, let alone these corrections, in the era of Advanced LIGO and Virgo.

\item Comparison between Penner et al. and our analysis: 
\begin{figure}[ht]
\includegraphics[width=0.5\textwidth]{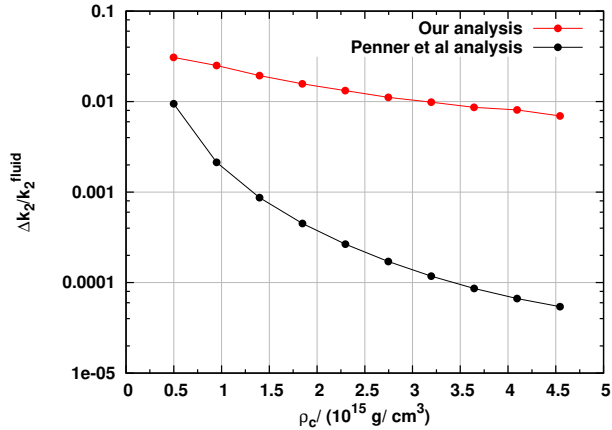}\\
\caption{Comparison between Penner et al.~\cite{Penner2011} and our analysis.}
\label{fig:delk_comparison}
\end{figure} 
Finally, we compare the results of our analysis with that of Penner et al. \cite{Penner2011}.  In their analysis they used a polytropic model for EOS and a simple profile of shear modulus that varies linearly with pressure. We implement the same EOS and shear modulus profile in our analysis and compare the obtained results with that of Pennal et al. in Fig. \ref{fig:delk_comparison}. We see that the changes in $k_2$ for our calculation is up to two orders of magnitude larger than their analysis. However, the change in $k_2$ is still small and the effect of solid crust in $k_2$ is unlikely to be observed by LIGO and Virgo detectors in the near future. This may, however, change in the detectors of the subsequent generation.   
%We use the numerical results of Table I of Penner et al.'s paper~\cite{Penner2011} to reproduce the change in $k_2$ as a function of central density same as Penner et al did.
\end{itemize}

\section{Conclusion}
\label{sec:conclusion}

In this paper, we have investigated the effect of elastic crust on the tidal deformation of neutron stars. We presented a complete set of static perturbation equations both for fluid core and solid crust of a neutron star. We verified here that our static perturbation equations are consistent with the zero frequency limit of Finn~\cite{Finn1990} but at  variance with some of the expressions in Penner et al. \cite{Penner2011}.
Recent independent calculation by Lau et al. \cite{Lau2018} have also pointed out that the set of static perturbation equations given by Penner et al. \cite{Penner2011} are inconsistent with zero frequency limit of pulsation equations of Finn~\cite{Finn1990}.
Our paper should be seen as an extension of the work done by Penner et al. \cite{Penner2011} who used a simple model of neutron stars based on a polytropic EOS and a simple linear profile for crustal shear modulus. In this paper, we have investigated the effect of realistic EOSs and realistic model of shear modulus on the tidal deformability of neutron stars. We find that realistic EOSs and shear modulus can cause a  change of $\sim 1\%$ in electric tidal Love number, much
larger than found by Penner et al. While this change may not be of much consequence for LIGO-Virgo observations in the near future, they may be important for subsequent generations of detectors.
%\section{Tentative Plan of the Paper}
%The paper can have two major motivations.
%1.) We will start with showing the effect of different crust EOS on $\lambda$. We will use the same core EOS and different crust EOS, consistently joined by Rana. Example. GM1 core and skm, sly4, sk272 crust. We can also take other variants of core and crust EOS. But, here we will not consider the shear. Rana will be able to provide more EOS.  
%2.) Bhaskar had modified Andersson's formulation for Love number calculation with shear following Finn's prescription. In this part, we will study the effect of shear modulus and compare with the fluid results. Bhaskar already has some preliminary results which he will upload in the work log. 
%Comments and suggestions are welcome. 

%CHECK: \sukanta{Address: how much does this change $\Lambda$? Radius?}
%\pc{The radius should not change as the background is unstrained. Therefore, the change in $\Lambda$ should be similar to the change in the Love numbers.}

\section*{Acknowledgements}

We would like to thank Sayak Datta and Kabir Chakravarti for helpful discussions. We also thank Wolfgang Kastaun for carefully reading the manuscript and making useful comments. This work was supported in part by the Navajbai Ratan Tata Trust. Bhaskar Biswas acknowledges the University Grant Commission (UGC) India, for the financial support as a senior research fellow.

\section*{Appendix: Dimensionless form} 
%\appendix*
%\section{Dimensionless form}
\label{appendix}
\subsection{Background equations}
 We introduce following dimensionless variables:
 \begin{equation}
     \rho=\rho_c\Tilde{\rho},\quad p=p_c\Tilde{p},\quad r=r_0 x\quad {\rm and}\quad m=m_0\Tilde{m}\, ,
 \end{equation}
 where $\rho_c$ and $p_c$ are central density and pressure, respectively. So, the TOV equations take the form
 \begin{eqnarray}
 \frac{d\Tilde{m}}{dx} &=& x^2\Tilde{\rho} \\
 \frac{d\Tilde{p}}{dx} &=& \frac{(\Tilde{\rho} + b\Tilde{p})(\Tilde{m}+x^3 b \Tilde{p})}{x(x-2b\Tilde{m} )}\\
 \frac{d\nu}{dx} &=& -\frac{b}{(\Tilde{\rho}+b\Tilde{p})}\frac{d\Tilde{p}}{dx},
\end{eqnarray} 
with
\begin{eqnarray}
b & = & \frac{p_c}{\rho_c}\\
m_0&=&4\pi r_0^3\rho_c \\
r_0^2 &=& \frac{b}{4\pi\rho_c}
\end{eqnarray}

\subsection{Perturbation equations for even parity}

Here we introduce additional dimensionless variables as:
\begin{eqnarray}
\Tilde{V} = \frac{V}{r^2},\quad \Tilde{W} = \frac{W}{r^3},\quad \mu = p_c \Tilde{\mu}.
\end{eqnarray}
Now we get following equations ($'=d/dx$):
\begin{eqnarray}
\frac{d\Tilde{V}}{dx} &=& e^{\lambda/2}\frac{\Tilde{W}}{x} - \frac{\Tilde{B}}{\Tilde{\mu}}\frac{e^\lambda}{x^2}\\
\Tilde{B} &=& \frac{B}{r_0p_c}\\
\frac{d\Tilde{W}}{dx} &=& \frac{e^{\lambda/2}}{x}\left[\frac{3\Tilde{A}}{4\Tilde{\mu}} - \frac{1}{2}(K - H_0)
           + \left( 16\pi d \Tilde{\mu}x^2 + 3\right)\Tilde{V}\right]\\
\Tilde{A}  &=& \frac{A}{p_c},\quad  d=p_c r_0^2\\
\delta \Tilde{p} &=&\frac{\delta p}{p_c} = \frac{\Tilde{\rho}+b\Tilde{p}}{b}c_s^2 \left[-\frac{3\Tilde{A}}{4\Tilde{\mu}}+ \frac{3}{2}K - 9\Tilde{V} + e^{-\lambda/2}\left(-3+\frac{x\nu'}{2c_s^2}\right)\Tilde{W}\right]\\
4b^2(\delta\Tilde{p}-\Tilde{A})x^2 &=& 4e^\lambda K - H_0(6e^\lambda -2 +x^2\nu'^2) - x^2\nu'H_0' - 4b^2\Tilde{\mu}\Tilde{V}x^4\nu'^2 + 4b^2e^\lambda x \Tilde{B}(2+x\nu')\\
\frac{dK}{dx} &=& H_0\nu' + H_0' + 4b^2\Tilde{\mu}(x\nu' +2 )x\Tilde{V} - 4b^2\Tilde{B}e^\lambda\\
\frac{d\Tilde{B}}{dx} &=& \frac{e^{-\lambda}}{4b^2x}(\nu' + \lambda')H_0 - \frac{\Tilde{B}}{2x}( 4 + x\lambda' + x \nu') -4\Tilde{\mu}\Tilde{V} + \delta\Tilde{p} + \frac{\Tilde{A}}{2}
\end{eqnarray}
\begin{eqnarray}
    -x^2H_0'' &+& \left[\frac{1}{2}x(\lambda'-\nu')-2\right]xH_0' + \left[6e^\lambda + 2(e^\lambda-1)-x(\lambda'+3\nu') + x^2\nu'^2\right]H_0 \nonumber\\
    &=& 2b^2x^2\Bigg\{-e^{\lambda}\delta\Tilde{p}(3+c_S^{-2})+8\Tilde{\mu}\Tilde{V}\left[1-e^{\lambda}+x\left(\nu'+\frac{1}{2}\nu'\right)-\frac{1}{4}x^2
    \nu'^2\right]\nonumber \\ 
    &+&4x\nu'\Tilde{\mu}\Tilde{V} + 2x^2\nu'(\Tilde{\mu}\Tilde{V})'+2\nu'\Tilde{B}e^{\lambda}\Bigg\}
\end{eqnarray}
Boundary conditions at the center (for $l=2$) become :
\begin{eqnarray}
    H_0 &=& ax^2 \\
    K &=& ax^2 \\
    \Tilde{V} &=& c \\
    \Tilde{W} &=& -2c,
\end{eqnarray}
Where $a$ and $c$ are constants.
Interface conditions are :
\begin{eqnarray}
    \Tilde{A_i} &=&\delta\Tilde{P_i}-\delta\Tilde{P_f}\\
    \delta\Tilde{p_f}&=&\frac{1}{2}\frac{\Tilde{\rho}+b\Tilde{p}}{b}H_{of}\\
    \delta\Tilde{p_i} &=& \frac{\Tilde{\rho+b\Tilde{p}}}{b}c_S^2
    \Bigg[-\frac{3\Tilde{A}}{4\Tilde{\mu}} + \frac{3}{2}K_i -9\Tilde{V} + e^{\lambda/2}\Tilde{W}\left(-3+\frac{x\nu'}{2c_S^2}\right) \Bigg]
\end{eqnarray}

\subsection{Perturbation equations for odd parity}

Dimensionless form of odd parity perturbed equation is:
 \begin{equation}
	    h_0^{''}-\frac{\lambda^{'}+\nu^{'}}{2}h_0^{'}+\left[\frac{\lambda^{'}+\nu^{'}}{x}-\frac{4e^{\lambda}}{x^{2}}-\frac{2}{x^{2}}+16\pi d \Tilde{\mu} e^{\lambda}\right]h_0=0
	        \end{equation}

\end{document}